\documentstyle[12pt]{article}
\def\lvm{\leavevmode\hbox to\parindent{\hfill}}

\def\o{\omega}
\def\l{\lambda}
\def\d{\partial}
\def\Det{\mathop{{\rm Det}}\nolimits}
\def\det{\mathop{{\rm det}}\limits}
\def\b{\beta}
\def\g{\gamma}
\def\ph{\phantom}
\def\noi{\noindent}
\def\req#1{(\ref{#1})}
\def\bF{{\bar\varphi}}
\def\vF{{\varphi}}
\def\cD{{\cal D}}
\def\cI{{\cal I}}
\def\cN{{\cal N}}
\def\hH{{\hat H}}
\def\cL{{\cal L}}
\def\ket#1{{\bigl|#1\bigr\rangle}}
\def\BE{\begin{equation}}
\def\EE{\end{equation}}
\def\BA{\begin{array}}
\def\EA{\end{array}} 

\makeatletter
\@addtoreset{equation}{section}
\newdimen\normalarrayskip
\newdimen\minarrayskip
\normalarrayskip\baselineskip
\minarrayskip\jot
\newif\ifold \oldtrue \def\new{\oldfalse}
\def\arraymode{\ifold\relax\else\displaystyle\fi}

\def\@arrayskip{\ifold\baselineskip\z@\lineskip\z@
  \else
  \baselineskip\minarrayskip\lineskip2\minarrayskip\fi}
\def\@arrayclassz{\ifcase \@lastchclass \@acolampacol \or
\@ampacol \or \or \or \@addamp \or
 \@acolampacol \or \@firstampfalse \@acol \fi
\edef\@preamble{\@preamble
 \ifcase \@chnum
  \hfil$\relax\arraymode\@sharp$\hfil
  \or $\relax\arraymode\@sharp$\hfil
  \or \hfil$\relax\arraymode\@sharp$\fi}}
\def\@array[#1]#2{\setbox\@arstrutbox=\hbox{\vrule
  height\arraystretch \ht\strutbox
  depth\arraystretch \dp\strutbox
  width\z@}\@mkpream{#2}\edef\@preamble{\halign \noexpand\@halignto
\bgroup \tabskip\z@ \@arstrut \@preamble \tabskip\z@ \cr}%
\let\@startpbox\@@startpbox \let\@endpbox\@@endpbox
 \if #1t\vtop \else \if#1b\vbox \else \vcenter \fi\fi
 \bgroup \let\par\relax
 \let\@sharp##\let\protect\relax
 \@arrayskip\@preamble}
\makeatother

\begin{document}

\large
\thispagestyle{empty}
\begin{center}
{\Large{\sc The Problem of Classical Limit in\\[2pt]
Quantum Cosmology:\\[8pt]
the Effective Action Language
\footnote{
Report at the Seminar ``Quantum Gravity. In Memory of Academician
Moisei Markov,'' June 12--19, 1995, Moscow.
Presented by B.~L.~Altshuler}
\footnote{
 E-mail: \tt altshul@td.lpi.ac.ru}}}\\[20pt]

\vspace{3ex}

{ \large B.~L.~Altshuler, A.~M.~Boyarsky and
A.~Yu.~Neronov}

{\large Department of Theoretical Physics, P.~N.~Lebedev Physical
Institute,

  Leninsky prospect 53, 117924 Moscow, Russia}\vspace{5ex}

\end{center}

\centerline{{\Large\bf Abstract}}

\normalsize

\begin{quote}
 The tool of functional averaging over some ``large''
diffeomorphisms is used to describe quantum systems with constraints,
in particular quantum cosmology, in the language of quantum Effective
Action.  Simple toy models demonstrate a supposedly general
phenomenon: the presence of a constraint results in ``quantum repel''
from the classical mass shell.
\end{quote}

\newpage

\addtolength{\baselineskip}{3pt}

\section{Introduction}
   At this memorial Quantum Gravity Conference dedicated to Moisei
Alexandrovich Markov I~think we must also remember some others who
unfortunately passed away: Andrei Sakharov, Yuri Golfand, Yakov
Zeldovich, and Igor Tamm whose centenary is being celebrated these days.
Professor Stelle has already presented here some reminiscences on
the political troubles concerning Sakharov and Golfand.  I would like
just to remind the audience that after highest political tops of the Former
USSR had reacted to the strongest international pressure and restored Yuri
Golfand in the Lebedev Institute, he, from 1980 and almost 10 years to the
moment of his departure to Israel, worked in the laboratory headed by Moisei
Alexandrovich Markov.

\bigskip

Classical limit in quantum cosmology becomes a problem because of the
Wheeler--De~Witt (WDW) and shift constraints:
\begin{eqnarray}
\hH\ket\Psi&=&0\,,\label{(1-1)}\\
\hH_i\ket\Psi&=&0\,,\label{(1-2)}
\end{eqnarray}

An enormous number of papers on the topic come to a more or less unique
conclusion: the construction of time and of the classical space-time is not
possible from the pure wave function of the Universe; the system must be
open, some external environment is necessary to provide decoherence, i.e.\ a
wave function reduction from the time-independent WDW state $\ket\Psi$ to the
states localized in time.  For the quantum-field theorist this just means
that gauge freedom responsible for constraints~\req{(1-1)}, \req{(1-2)} is
broken extrinsically.  Another way to eliminate the problem is to use unitary
gauge (see e.g.\ review \cite{1}), i.e.\ to nominate one of the quantum
coordinates (e.g.\  scale factor of the Universe, or some scalar field) to be
a nonquantized classical clock.  One can also use Mensky channels \cite{2},
i.e.\ to make a substitution (symbolically):
\BE
\delta(H)=\int e^{i N H}dN\Rightarrow\int e^{-\gamma N^2}e^{i N H}dN\, .
\label{(1-3)}
\EE

An artificial weight factor for the Lagrange multiplier $N$ in \req{(1-3)},
nonquantized clocks, `environment' --- in all these approaches the question
about the classical limit in quantum cosmology is in a sense erased `by
hand'. Here we consider the gauge freedom of general coordinate
transformations (and also the global `residual gauge freedom' -- see Sec.~2)
seriously; it will be shown that in the Effective Action (EA) language, the
presence of constraints results in the `quantum repel' from the mass
shell, i.e.\ "repel" from the solutions of the classical dynamical equations
\BE
{\delta S\over\delta\varphi^A}=0\,,
\label{(1-4)}
\EE

\noindent
where $S$ is the classical Action.  Only one-loop corrections to EA will be
considered:
\BE
\Gamma(\bF)=S(\bF)+\Gamma^{(1)}\Bigl(\bF,\displaystyle{\delta S \over
{\delta\bF^A}}
\Bigr)\, .
\label{(1-5)}
\EE

\noindent
$\bF$ are the background fields.  Quantum repel from the mass shell
means that because of the constraints the one-loop term $\Gamma^{(1)}$
possesses a specific nonlocal dependence on
${{\delta S}\over{\delta\bF^A}}$ and is nonanalytic or divergent on the
solutions of Eq.~\req{(1-4)}.  `Bad' behaviour of the one-loop term near $\bF$
means the divergence of quantum dispersion and
signals that this background is not a classical field.  New mass shell
(and in particular a regularized vacuum $\bF_0$) is defined as a
solution of the equation
\BE
{\delta\Gamma\over\delta\bF^A}=0
\label{(1-6)}\EE

\noindent
in full analogy with the standard procedure, e.g.\ in the Coleman--Weinberg
model~\cite{3}.  The `external current' $\displaystyle{\delta
S\over\delta\bF_0}$ may
be found self-consistently from~\req{(1-5)} and \req{(1-6)}.  (In this
way dynamically generated mass of the Goldstone boson may be found in the
Glashow-Higgs model \cite{4}).

\section{Constraints and functional integration}
It is well known that restriction of the Hilbert space to some subspace
results in nontrivial kinematics and dynamics of the physical observables and
also results in problems with the classical limit \cite{5}--\cite{7}.  E.g.,
if we postulate that only even states $\psi(x)=\psi(-x)$ of the most popular
one-dimensional oscillator are permitted, then not $x$ but $|x|$ becomes a
relevant physical observable and contrary to the conventional oscillator
there is no classical limit near $x=0$~\cite{5}. The same is true for the
dynamics of the radial coordinate of a free particle constrained to the
$S$-state of its Hilbert space~\cite{8}.  Van~der~Monde determinant and the
phenomenon of quantum `level repel' repeat the story for the unitary matrices
averaged over angles~\cite{9}.

Constraints -- functional integration over Lagrange multipliers -- gauge
theory and gauge fixing -- Faddeev--Popov ghosts -- Generating Functional and
the Effective Action: this well known logical chain puts the following
questions:
\begin{enumerate}
\item How to construct a theory whose physical predictions do not depend on the
gauge fixing?
\item What is left of a constraint after gauge-fixing?
\end{enumerate}

As an answer to the first question we use here the Vilkovisky--De$\,$Witt
unique effective Action where relativistic Landau--De$\,$Witt, the so called
`natural', gauge is used (see Sec.~3).

Let us discuss the second question?  If the gauge freedom is over-fixed (as
in RHS of \req{(1-3)}) then physics of the constraint is lost.  If we
conventionally postulate that functional integration is done over fields
(including gauge diffeomorphisms) which are zero at infinity (at the
boundary) then again the constraint is lost in the Green functions Generating
functional and in EA; it must be imposed in the asymptotic Hilbert space as a
supplementary condition.  Instead of it one may generalize the definition of
the functional integral so as to include into it the integration over some
`global' gauge transformations, nonzero at the infinity . We shell go this
way.

This approach is not a new one.  In quantum cosmology Teitelboim~\cite{10}
and Halliwell~\cite{11} showed that WDW constraint \req{(1-1)} follows from
the integration over a residual gauge mode $N=$const (which corresponds to
time reparametization $t\to t+{\rm const} \cdot t$) i.e.\ over Faddeev--Popov
zero (`residual gauge') in the $\delta({\dot\cN})$ gauge fixing.  The
supplementary integration over global gauge transformations is equivalent to
the requirement of wave function invariance under `large' diffeomorphisms and
physically means that some charges are zero~\cite{12} -- and these are
constraints. The residual gauge contribution to the functional integral is
responsible for the specific IR peculiarities of EA at the mass
shell~\req{(1-4)}; their regularization by external currents follows from the
general formula~\req{(3-19)} below (Cf.~\cite{13}).

\section{Vilkovisky--De~Witt Effective Action (VDWEA)}
Conventional EA is defined through generating functional $W(j)$
\cite{14}:\nopagebreak
\BE
\Gamma(\bF)=W(j)-\bF^Aj_A\,,
\label{(3-1)}
\EE

\noindent
where $\bF^A=\displaystyle{{\delta W}\over{\delta j^A}}$, ~~$A$ is a condensed
\hbox{De$\,$Witt}
index (continuous and discrete), and
\BE
e^{iW(j)}=\int\cD\varphi e^{iS(\varphi)+i\varphi^Aj_A}\,.
\label{(3-2)}
\EE

\noindent
However this EA suffers from two ambiguities: \ (1)~it depends on the local
fields
redefinition~\cite{15}; \ (2) in gauge theories EA depends (away from the mass
shell)
on the gauge fixing term.

The VDWEA approach \cite{16,17} (it is also
applied to gravity in~\cite{18} and to scalar QED in~\cite{19}) heals the
first disease by considering functional space $\varphi^A$ as a Riemannian space
with an interval:
\BE
d\ell^2=d\varphi^AG_{AB}(\varphi)d\varphi^B\,.
\label{(3-3)}
\EE

\noindent
In particular, for the gravitational field ($\varphi^A=g_{\mu\nu}$)
\BE
G_{AB}=\sqrt{-g}\,G^{\alpha\beta\gamma\delta}
\delta(x-x')={1\over 2}\sqrt{-g}\,(g^{\alpha\gamma}g^{\beta\delta}
+g^{\alpha\delta}g^{\beta\gamma}-cg^{\alpha\beta}g^{\gamma\delta})
\delta(x-x')
\label{(3-4)}
\EE

\noindent
($c$ is an arbitrary constant).

Now instead of the ordinary functional derivatives, functionally covariant
derivatives are used, e.g.
\BE
S_{,AB}\to S_{;AB}=S_{AB}-\left\{{C\atop AB}\right\}S_{,C}
\label{(3-5)}
\EE

\noindent
where Cristoffel symbols are built from the metric~\req{(3-3)}. Hence loops
do not depend on the choice of coordinate in the functional space, i.e.\ on
the fields reparametrization.

In gauge theories, the generators $K^A_\gamma(\vF)$ of
the gauge transformations $\xi^\gamma$
\BE
\delta\vF=K^A_\gamma(\vF)\xi^\gamma\equiv\eta^A_{(\xi)}
\label{(3-6)}
\EE

\noindent
(parametrized by the fields $\xi^\gamma$) must be Killing vectors of the
metric~\req{(3-3)}. In particular in gravity we have
\BE
\delta\vF^{(g_{\mu\nu})}\equiv\delta g_{\mu\nu}=
-\xi_{\mu;\nu}-\xi_{\nu;\mu}=
(-g_{\mu\gamma}\nabla_\nu-g_{\nu\gamma}\nabla_\mu)
\xi^\gamma =
K_\gamma^{(\mu \nu)} \xi^\gamma
\equiv\eta^{(\mu\nu)}_{(\xi)}\,.
\label{(3-6)'}
\EE

At every `world' point $\bF^A$ of the functional space we can take
differentials $\eta^A$ of the field:
\BE
\vF^A=\bF^A+\eta^A
\label{(3-7)}
\EE

\noi
in the direction normal to the gauge orbit:
\BE
K^A_\gamma G_{AB}\eta^B\equiv(K_\gamma\eta)=0 \,.
\label{(3-8)}
\EE

\noi
This is the Landau--\hbox{De$\,$Witt}, or the so called ``natural", gauge
condition:
e.g.\ $A^\mu_{\ph{\mu},\mu}=0$ in QED, and
\BE
\chi_\gamma\equiv(h^\mu_\gamma-{c \over 2}\delta^\mu_\gamma h_\nu^\nu)_{;\mu}
=0
\label{(3-8')}
\EE

\noi
for free gravity ($g_{\mu\nu}=\bar g_{\mu\nu}+ h_{\mu\nu}$) with the functional
metric~\req{(3-4)}.   The substitution
$\eta^A=\eta^A_{(\xi)}$ from \req{(3-6)} to \req{(3-8)} gives a Faddeev--Popov
differential operator
\BE
Q_{\gamma\beta}=K^A_\g G_{AB}K^B_\beta
\label{(3-9)}
\EE

\noi
and equation for the "residual" gauge transformations
$\xi_0^\beta$:
\BE
Q_{\g\b}\xi^\b_0=0\,;
\label{(3-10)}
\EE

\noi
in free electrodynamics~\req{(3-10)} is $\Box\xi_0=0$ ($\Box$ is a
D'Alambertian), and in free gravity it is (from~\req{(3-6)'}, \req{(3-8')}):
\BE
(g_{\mu\nu}\Box+\nabla_\nu\nabla_\mu-c\nabla_\mu\nabla_\nu)\xi^\nu_0=0\,.
\label{(3-10')}
\EE

In the most simplified formulation, the VDWEA remedy against the second
ambiguity of EA mentioned at the beginning of this Section is: ``Use
`natural' gauge-fixing~\req{(3-8)}''. Actually this is true only for one-loop
terms; special Vilkovisky connection~\cite{17} must be introduced to
calculate unique VDWEA in higher orders in $\hbar$.  But in this paper we are
interested only in the one-loop term $\Gamma^{(1)}$.  Thus
\BE
e^{i\Gamma^{(1)}}=\lim_{\alpha\to\infty}\int e^{iS^{(2)}}
\Det \widehat Q ~ M{ \left[ D\eta^A \right]}\,.
\label{(3-11)}
\EE

\noi
Here $S^{(2)}$ is a sum of a second functionally covariant variation of the
classical Action $S$ and the gauge-fixing term:
\BE
S^{(2)}(\eta) \equiv {1 \over
2}\eta^A\stackrel{\leftrightarrow}{F}^{}_{AB}\eta^B
={1 \over 2}\eta^AS^{}_{;AB}\eta^B+{\alpha \over 2}(K_\g\eta)f^{\g\delta}
(K^{}_\delta\eta)\,;
\label{(3-12)}
\EE

\noi
$f^{\g\delta}$ is some arbitrary matrix in the space of gauge indices; the
measure reads
\BE
M=[\Det(\alpha f^{\g\delta})]^{1/2}\,;
\label{(3-13)}
\EE

\noi
the other symbols are defined in \req{(3-5)}, \req{(3-7)}, and \req{(3-8)}.
Finally we have VDWEA to the one-loop term:
\BE
\Gamma(\bF)=S+\Gamma^{(1)}=S+{1 \over 2}\lim_{\alpha\to\infty}
\left\{\log\Det(D^{AB}G_{BC})+
\log M\right\} +\log\Det \widehat Q\,,
\label{(3-14)}
\EE

\noi
where
\BE
D^{AB}=(\vec{F}_{AB})^{-1}
\label{(3-15)}
\EE

\noi
and the differential operator $\vec{F}_{AB}$ is defined in \req{(3-12)};
$G_{AB}$, $\widehat Q$, and $M$ are defined in (3.3), (3.11) and (3.16).

The identify
\BE
S_{,B}K^B_\g\xi^\g=0
\label{(3-16)}
\EE

\noi
represents gauge-invariance of the Action $S$.  Its covariant functional
differentiation gives
\BE
S_{;AB}K^B_\g\xi^\g=-S_{,B}K^B_{\g;A}\xi^\g \,.
\label{(3-17)}
\EE

\noi
Thus for the `residual gauge' field variations
\BE
\eta^A_0=K^A_\g\xi^\gamma_0
\label{(3-18)}
\EE

\noi
(where  $\xi^\gamma_0$ are solutions to~\req{(3-10)}) we have
from~\req{(3-12)},
with account of (3.9) and (3.20):
\BE
\vec{F}_{AB}\eta^B_0=-S_{,B}K^B_{\g;A}\xi^\g_0\,,
\label{(3-19)}
\EE

\noi
hence $\eta^A_0$ are zeros of the differential operator $\vec F$ on the mass
shell
of the background field, i.e. when $S_{,B}\equiv\delta S/\delta\bF^B=0$.
Zero modes \req{(3-18)} are conventionally excluded from the definition of
$D=F^{-1}$ in \req{(3-14)} (i.e.\ from the functional integration in
\req{(3-11)}) because they are nonvanishing all over the space-time including
the infinity.  \underline{Our hypothesis:} they must be taken into account if
we want to
preserve constraints' physics in the functional integral language.

\medskip

\noi
{\sc Note}.
In General Relativity (GR), contrary to the vector gauge theories, the RHS of
the
identity~\req{(3-16)} is not zero but is a surface integral: under general
coordinate transformation $x^\mu\to x^\mu+\xi^\mu(x)$, we have
\BE
\delta S=\delta\left(\int\cL\sqrt{-g}\,dx\right)=
\int(\cL\sqrt{-g}\,\xi^\mu)_{,\mu}dx\,.
\label{(3-20)}
\EE

\noi
Thus $S^{(2)}(\eta_0)={1\over 2}\eta^A_0
\stackrel{\leftrightarrow}{F}_{AB}\eta^B_0$ (with $\eta_0$ taken from
\req{(3-18)}) is zero on the mass shell only for those solutions to
Eq.~\req{(1-4)} which render gravitational Lagrangian $\cL=0$ at the boundary.
Nevertheless \req{(3-19)} is valid in GR as
well as in other gauge theories.

\section{Simple quantum-mechanical example}
Before turning to cosmology, let us consider the simplest illustrative
theory: scalar QED reduced to quantum mechanics.  Its Action and the
functional space interval \req{(3-3)}:
\begin{eqnarray}
S&=&\int\left[{1\over 2} \dot \rho^2+{1\over 2} \rho^2(\dot x -
ey)^2-V(\rho)\right]dt
\label{(4-1)}\\
d\ell^2&=&[d\rho^2+\rho^2dx^2-dy^2]dt
\label{(4-2)}
\end{eqnarray}

\noi
are the reduced version of the Glashow--Higgs model (the scalar field
$\vF=\rho e^{ix}$; $x$ is the Goldstone field; $A_\mu=\{A_0,0,0,0\}$;
$A_0\equiv y(t)$; $e$ is the electric charge, and the dot symbolizes $d/dt$;
in~\req{(4-1)} and \req{(4-2)} all quantities are normalized to be
dimensionless). $V(\rho)$ is a standard Higgs potential
\BE
V={{m^2}\over {8\rho_0^2}}(\rho^2-\rho_0^2)^2\,.
\label{(4-3)}
\EE

\noi
($m$ is the mass of the $\rho$-field at the extremum of $V$). We shall
quantize $x(t),y(t)$ and not $\rho(t)$, and consider VDWEA correction to the
potential (4.3). To calculate the effective potential
\BE
V_{\rm eff}=-{1 \over \theta}\Gamma=V+V^{(1)}
\label{(4-4)}
\EE

\noi
(with $\theta$ being the time interval) we conventionally take a background
\BE
\bar\rho(t)=\rho={\rm const}\,,\qquad\bar x(t)=\bar y(t)=0\,.
\label{(4-5)}
\EE

\noi
Now $x,y$ may be considered as small variations
\BE
\eta^A=\{x;y\}\,,
\label{(4-6)}
\EE

\noi
their gauge transformations
\BE
\delta x=K^{(x)}\xi=e\xi\,;\qquad
\delta y=K^{(y)}\xi=\dot\xi
\label{(4-7)}
\EE

\noi
leave~\req{(4-1)} invariant. The natural gauge~\req{(3-8)} and the
Faddeev--Popov equation~\req{(3-10)} take the form (we used the metric
\req{(4-2)}:
\begin{eqnarray}
&&\stackrel{\leftarrow}{{d\over dt}}G_{yy}\cdot y +
eG_{xx}\cdot x=
-\stackrel{\rightarrow}{{d\over dt}}
G_{yy} y +
eG_{xx}\cdot x=
\dot y+e\rho^2x=0\,,
\label{(4-8)}\\
&&Q\xi_0=\stackrel{\cdot\cdot}{\xi}_0+e^2\rho^2\xi_0=0\,.
\label{(4-9)}
\end{eqnarray}

Because of the dependence of the $G_{xx}$ component of the metric \req{(4-2)}
on $\rho$, \ $S_{;AB}$ in~\req{(3-12)} includes the mass term of the
``Goldstone boson" $x(t)$ (see~\req{(3-5)}):
\BE
\left\{{\rho\atop\b\b}\right\}S_{,\rho}=
-\rho{dV\over d\rho}\equiv-\rho^2\mu^2\,,
\label{(4-10)}
\EE

\noi
which, as is well known, is zero on the mass shell, i.e.\  at the extremum of
the potential \req{(4-3)}. From~\req{(4-1)}, \req{(4-8)}, \req{(4-10)} we
have for this model the Action \req{(3-12)}:
\begin{eqnarray}
S^{(2)}(\eta)&=&\int\left\{
{1 \over 2}\rho^2(\dot x-ey)^2-{1 \over 2}\rho^2\mu^2x^2+
{\alpha \over 2}(\dot y + e\rho^2x)^2\right\}dt
\nonumber\\
{}&\equiv&{1 \over 2}
\eta^A\stackrel{\leftrightarrow}{F}_{AB}\eta^B\,.
\label{(4-11)}
\end{eqnarray}

Performing the Fourier transform we arrive at the following expression for
the determinant of the differential operator $\vec F$ in~\req{(4-11)} in the
$\{x,y\}$ linear space:
\BE
\det F(p)=\alpha Q^2-\mu^2(\alpha p^2+e^2\rho^2)
\equiv\alpha(p^2-m_1^2)(p^2-m_2^2)
\label{(4-12)}
\EE

\noi
where
\BE
Q(p)=p^2-e^2\rho^2
\label{(4-13)}
\EE

\noi is the Fourier transform of the Faddeev--Popov operator \req{(4-9)};
\BE
m_{1,2}^2=e^2\rho^2+{1 \over 2}\mu^2\pm{1 \over 2}\sqrt{
\mu^4+4e^2\mu^2\rho^2(1+\alpha^{-1})}\,\,.
\label{(4-14)}
\EE

Substitution of ~\req{(4-12)}, \req{(4-13)}, and $\prod\alpha^{1/2}$
\req{(3-13)} into~\req{(3-14)} gives the one-loop term of the EA.  To
calculate it we first take its derivative (see dependence on $\mu^2$
in~\req{(4-12)}):
\BE
{d\Gamma^{(1)}\over d\mu^2}=
{\theta i\over4\pi}\int{p^2+e^2\rho^2\alpha^{-1}\over
(p^2-m_1^2)(p^2-m_2^2)-i\epsilon}dp=
-{\theta \over4(m_1-m_2)}\left(1+{e^2\rho^2\alpha^{-1}\over
m_1m_2}\right)\,.
\label{(4-15)}
\EE

Zeroes of the denominator in the integrand in~\req{(4-15)} are located in
the upper half-plane at $p=m_1$ and $p=-m_2$.  Because of this fact
\req{(4-15)} is proportional to
\BE
(\cD^c_{m_1}-\bar\cD^c_{m_2})\Bigm|_{t=0}
\label{(4-16)}
\EE

\noi ($\cD^c$, $\bar\cD^c$ are causal and anticausal Green functions), that
is why \req{(4-15)} is divergent when $m_1\to m_2$, i.e.\  at the extremum of
$V(\rho)$ where $\mu=0$. From~\req{(4-15)}, \req{(4-4)} we finally have (for
$\alpha=\infty$):
\BE
V_{\rm eff}(\rho)=V(\rho)-\int{d\mu^2\over4(m_1-m_2)}\;.
\label{(4-17)}
\EE

\noi
For $\mu\ll\rho$ this gives, with the account of ~\req{(4-10)}, \req{(4-14)}:
\BE
V_{\rm eff}(\rho)\approx V(\rho)-{1\over 2}\sqrt{{{1\over\rho}{dV\over
d\rho}}}\,.
\label{(4-18)}
\EE

\noi
(The `charge' $e$ does not enter the answer in this approximation.  In
\cite{4} we considered  Glashow--Higgs model in 4 dimensions.
There, the Goldstone boson mass $\mu$ at the extremum of $V_{\rm eff}(\rho)$
depends on $e,\rho_0,m$). Formula~\req{(4-18)} is the main result of this
Section and also illustrates the main idea of the article: quantum repel from
the mass shell. Indeed the minimum $V_{\rm eff}$ is shifted away from the
minimum of the potential~\req{(4-3)}.

Let us contemplate a bit on this result. Substitution $\rho={\rm const}$ in
\req{(4-1)} makes this model the well known [20] simplest gauge theory
generated by the constraint
\BE
\hat p_x\ket\psi=-i{\d\over\d x}\ket\psi=0\,.
\label{(4-19)}
\EE

\noi
Eq.~\req{(4-19)} means that the scalar field $\vF=\rho e^{ix}$ is in the
$S$-state with regard to the phase $x$.  There is a standard ``philosophical"
question addressed to all theories with a spontaneous breakdown of symmetry:
``Who selects a definite phase of the order parameter in the `broken' vacuum
state?" Bogolyubov's quasi-averages are the most popular answer, but they
suppose the temporary (before $V^{(3)}\to\infty$) introduction of the
external field; this violates the constraint~\req{(4-19)} and prepares a
quantum state with a definite phase $x_0$. Here we postulate that the system
is permanently in the $S$-state~\req{(4-19)} and we come to the conclusion
that the minimum $\rho=\rho_0$ of the potential~\req{(4-3)} cannot be
realized in this case. The use of a noncausal contour in the longitudinal
sector (in the complex $p$-plain in~\req{(4-15)}) is crucial; supposedly it
is a counterpart of the constraint~\req{(4-19)} in the EA approach to the
theory~\req{(4-1)}.

Is this physically adequate?  To our mind this question is perhaps not too
adequate itself. Actually with the choice of the gauge-fixing and of the way
of averaging over `large' diffeomorphisms we do change the effective quantum
dynamics and in particular one of its the most important physical predictions
-- the effective `classical' dynamics, i.e.\ renormalized mass shell,
determined by the equation \req{(1-6)}. Thus postulating one or another way
of averaging over global gauge modes in the functional integral~\req{(3-11)}
we define the theory.

Our observation is that the presence of a constraint demands some sort of
global averaging.

\section{VDWEA language for the cosmological\hfill\break
minisuperspace model}
Earlier we investigated the VDWEA approach to the relativistic
particle~\cite{4}.  Now we consider the standard flat cosmological model in
$(n+1)$ dimensions with space-time metric
\BE
ds^2=e^{2\nu}dt^2-e^{2\b}d\Omega^{(n)}\,.
\label{(5-1)}
\EE

\noi
For this model, the reduced Einstein Action with a $\Lambda$-term and the
reduced functional space metric \req{(3-4)} are of the form
($L=V^{(n)}/2\kappa$, $V^{(n)}$ is the space volume, and $\kappa$ is the
gravitational constant in $n+1$ dimensions; in fact, $L$ is an arbitrary
length parameter of the reduced `flat' minisuperspace model ~\req{(5-2)}):
\begin{eqnarray}
S&=&\int\sqrt{-g}\,\cL^{(0)}dt=
L\cdot\int e^{\nu+n\b}\left[-n(n-1)e^{-2\nu}\dot\b^2-2\Lambda\right]dt\,,
\label{(5-2)}\\
d\ell^2&=&L^{-1}\int e^{\nu+n\b}[(2-c)d\nu^2-
2cnd\nu d\b-n(cn-2)d\b^2]dt\,.
\label{(5-3)}
\end{eqnarray}

The classical solution is the De-Sitter universe; in the proper time
($\nu=0$):
\BE
\b_{\rm cl}=\lambda_0t\,,\qquad\lambda_0^2={2\Lambda\over n(n-1)}\,.
\label{(5-4)}
\EE

\noi
We calculate one-loop VDWEA term \req{(3-11)} on the background:
\BE
\bar\nu=0\,,\qquad\bar\b=\lambda t\,,
\label{(5-5)}
\EE

\noi
$\l-\l_0 \neq 0$ is a measure of departure from the mass shell.  On the
background \req{(5-5)} we have the sources:
\BE\new\BA{rcl}
S_{,\nu}&=&Le^{n\l t}n(n-1)(\l^2-\l_0^2)\,,\\
S_{,\b}&=&nS_{,\nu}\,.
\EA
\label{(5-6)}
\EE

A reparametrization of time $t\rightarrow t + \xi(t)$ generates gauge
variations
which on the background \req{(5-5)} read
\BE
\delta\nu\equiv K^\nu\xi=\dot{\bar\nu}\xi+\dot\xi=\dot\xi\,,\qquad
\delta\b\equiv K^\b\xi=\dot{\bar\b}\xi=\l\xi\,.
\label{(5-7)}
\EE

Knowledge of the functional Killing vectors \req{(5-7)} and of the metric
\req{(5-3)} allows us to write down the Landau--\hbox{De$\,$Witt} gauge
condition
\req{(3-8)}
\BE
\chi\equiv(2-c)\dot\eta^\nu-cn\dot\eta^\b+2n\l(\eta^\nu-\eta^\b)=0\,,
\label{(5-8)}
\EE

\noi
and Eq.~\req{(3-10)} for Faddeev--Popov zero modes:
\BE
\widehat Q \xi_0 =
(2-c)\stackrel{\cdot\cdot}{\xi}_0+
\l n(2-c)\dot\xi_0-2\l^2n\xi_0=0\,.
\label{(5-9)}
\EE

\noi
Here,
\BE
\eta^A=\{\eta^\nu;\eta^\b\}\,,\qquad\eta^\nu\equiv d\nu\,,\quad
\eta^\b\equiv d\b\,.
\label{(5-10)}
\EE

Of course one may get formulae~\req{(5-8)}, \req{(5-9)} for the space-time
\req{(5-1)} directly from the general formulae~\req{(3-4)}, \req{(3-8')} and
\req{(3-10')}.

Omitting elementary calculations we present the final expression for
this model for $S^{(2)}$ from~\req{(3-12)}:
\BE\new\BA{rcl}
S^{(2)}&=&L\int e^{n\l t}\left[-n(n-1)(\dot\eta^\b-\l\eta^\nu)^2
-{{n}\over {n-1}}E(\eta^\nu-\eta^\beta)^2+{\alpha \over 2}\chi^2\right]dt\\
{}&{}&{}-L\int{d\over dt}\left[n^2(n-1)\l e^{n\l t}(\eta^\b)^2\right]dt\,,
\EA
\label{(5-11)}
\EE

\noi
where the gauge-fixing function $\chi$ is defined in \req{(5-8)}, and
\BE
E\equiv{n(n-1)^2\over cn+c-2}(\l^2-\l_0^2)\,.
\label{(5-12)}
\EE

\noi
{\sc Note}.
It is easily seen that the
Hessian $(\d^2\cL/\d\dot q^i\d\dot q^k)$ of the Action
$S^{(2)}$~\req{(5-11)} is
proportional to the local functional metric \req{(5-3)} if:
\BE
c=1\,,\qquad\alpha=1\,,
\label{(5-13)}
\EE

\noi
($\alpha=1$ is the so called Feynman gauge fixing).
This Canonical quantization self-consistent condition is valid on the proper
time background ($\bar\nu=0$ in \req{(5-1)}); and it is rather general
\cite{17}.  Following Vilkovisky, we put $c=1$ in the functional
gravitational metric \req{(3-4)}. But we leave $\alpha$ arbitrary and in the
end we put $\alpha\to\infty$, which gives the Landau--\hbox{De$\,$Witt} gauge
fixing \req{(5-8)}.

To pursue further calculations, the following substitution to
\req{(5-3)}, \req{(5-8)}, (5.11) is useful:
\BE
\eta^\b=e^{-{n\l t\over2}}x\,,\qquad
\eta^\nu=e^{-{n\l t\over2}}(y+nx)\,.
\label{(5-14)}
\EE

\noi
This diagonalizes functional metric~\req{(5-3)}:
\BE
d\ell^2=L^{-1}\int[-2n(n-1)x^2+y^2]dt\,,
\label{(5-15)}
\EE

\noi
makes the Faddeev--Popov differential operator~\req{(5-9)} self-conjugate,
\BE
\vec Q={d^2\over dt^2}-{1 \over 4} n(n+8)\l^2\,,
\label{(5-16)}
\EE

\noi
and results in the following expression for $S^{(2)}$ (5.11) (we
discarded surface terms, and put $c=1$ everywhere):
\BE\new\BA{rcl}
S^{(2)}&=&L\int \left[ -n(n-1)\dot x^2+{\alpha \over 2}\dot y^2
+2 n(n-1)\l(\dot xy+\alpha\dot yx)
+{1 \over 2} x^im_{ik}x^k \right] dt\\
{}&\equiv&\int{1 \over 2} x^i\stackrel{\leftrightarrow}{F}_{ik}x^kdt\,,
\EA\label{(5-17)}\EE

\noi
$x^i=\{x^1;x^2\}\equiv\{x;y\}$.  We shall not write down expressions for the
elements of the `mass' matrix $m_{ik}$, but present the final answer in the
momentum representation for the combination of determinats which, according to
general formula~\req{(3-14)}, determine $\Gamma^{(1)}$ ($i,k=\{1,2\}$):
\BE
\cI(p)\equiv {\det\vec F_{ik}\over
\det G_{ik}\cdot\det(\alpha\delta_{ik})(\det\vec Q)^2}=
{(p+\omega_1^2)(p+\omega_2^2)\over(p+\omega_{\rm F\mbox{-}P}^2)^2}\,,
\label{(5-18)}
\EE

\noi
$\vec F_{ik}$ is given by variations of $S^{(2)}$ in \req{(5-17)}:
\BE
{\delta S^{(2)}\over\delta x^i}=\vec F_{ik}x^k\,,
\label{(5-19)}
\EE

\noi
for $G_{ik}$ see~\req{(5-15)}, and for $\vec Q$,~\req{(5-16)};
\BE\new\BA{rcl}
\o_{1,2}^2&=&\o_0^2+{3 \over 4}~{{n+2}\over {n-1}}E\pm
\sqrt{{1 \over 4} ~{{n+7}\over{n-1}} E^2+2nE\l_0^2}\,,\\
\o_{\rm F\mbox{-}P}^2&=&\o_0^2+{1 \over 4}~ {{n+8}\over{n-1}}E\,,\\
\o_0^2&=&{1 \over 4} n(n+8)\l_0^2\,;\qquad E=n(n-1)(\l^2-\l_0^2)\,.
\EA\label{(5-20)}\EE

For the one-loop term of the effective Lagrangian (defined in \req{(5-27)}
below) we have finally:
\BE
\cL^{(1)}={1 \over 2\pi}\int^E\left[\int_C{d\log\cI\over dE}dp\right]\,dE\,.
\label{(5-21)}
\EE

\noi
The crucial point is the choice of the contour which selects two out of the
four poles of $\cI^{-1}$ \req{(5-18)}:  $p=i\o_1$ and $p=-i\o_2$ (see
discussion in Sec.~4).  With the choice of the noncausal pole at $p=-i\o_2$
we in a sense take into account the global Faddeev--Popov zeroes in the
functional integral.  With this choice of the contour in the $p$-plane and
with the account of the formula (see \req{(5-18)}, \req{(5-20)})
\BE\new\BA{rcl}
{d\log\cI\over dE}&=&
{Ap^2+BE+C\o_0^2\over
(p^2+\o_1^2)(p^2+\o_2^2)}
-{2D\over(p^2+\o_{\rm F\mbox{-}P}^2)}\,,\\
A&=&{3 \over 2}{{n{+}2}\over{n{-}1}}\,;\quad
B={{5n^2{+}12n{+}64}\over {8(n{-}1)^2}}\,;\quad
C={{3n^2{+}14n{+}64}\over{2(n{+}8)(n{-}1)}}\,;\quad
D={{n{+}8}\over{4(n{-}1)}}
\EA\label{(5-22)}
\EE

\noi
we have the following result of the momentum integration in \req{(5-21)}:
\BE
\int_C{d\log\cI\over dE}dp ={\pi\over\o_1-\o_2}
\left( A-{BE+C\o_0^2\over\o_1\o_2}\right)-
{2\pi D\over\o_{\rm F\mbox{-}P}}\,
\label{(5-23)}
\EE

\noi
(for $\o_1,\o_2,\o_0,\o_{\rm F\mbox{-}P},E$, see \req{(5-20)}, and for
$A,B,C,D$ see \req{(5-22)}).  The final integration over $E$ in \req{(5-21)}
is not elementary.  We present the answer for the one-loop term \req{(5-21)}
of the effective lagrangian for small $E$:
\BE
\cL^{(1)}=\sqrt{{8\over{n+8}}E}\,,\qquad E=n(n-1)(\l^2-\l_0^2)\ll\l_0^2\,.
\label{(5-24)}
\EE

\noi
(We normalized $\cL^{(1)}$  to zero at the mass shell $E=0$.) To see the
physics behind this
result let us remember that according to \req{(5-5)} the background parameter
$\l$ is the derivative of $\bar\beta$.  One may also change the background
value
of $\bar\nu$ from zero to any constant; this corresponds to the time
reparametrization $t\to {\rm const}\cdot t$ and according to \req{(5-7)} does
not
violate the crucial for our calculations condition
$\dot{\bar\b}={\rm const}$.
To write down gauge-invariant EA, the following substitution in the
expression for $\Gamma^{(1)}$ must be performed:
\BE
\l= \dot{\bar\b} \Rightarrow  e^{-\bar\nu}\dot{\bar\b}\,;
\qquad dt\Rightarrow e^{\bar\nu}dt\,;
\label{(5-25)}
\EE

Hence from the definition of $E$ in \req{(5-24)} and $\l_0^2$ in \req{(5-4)}:
\BE
E=n(n-1)e^{-2\bar\nu}\dot{\bar\b}^2-2\Lambda\,.
\label{(5-26)}
\EE

The Effective Action \req{(3-14)} is here a time integral of the sum of the
classical Lagrangian $\cL^{(0)}$ (see \req{(5-2)}) and the one-loop term
$\cL^{(1)}$ \req{(5-24)},
where $E$ is taken in the form \req{(5-26)}.  Finally:
\BE\new\BA{rcl}
\Gamma(\bar\nu,\bar\b)&=&\int\sqrt{-g}\,(\cL^{(0)}+\cL^{(1)})dt\\
{}&=&L\int e^{\bar\nu+n\bar\b}
\left\{\left[-n(n-1)e^{-2\bar\nu}\dot{\bar\b}^2-2\Lambda\right]
+L^{-1}\left({8\over {n+8}}\right)^{1/2}E^{1/2}\right\}dt\,.
\EA\label{(5-27)}
\EE

{}From (5.27) and (5.26) renormalized constraint equation
\BE
{\d\Gamma\over\d\bar\nu}=0\,.
\label{(5-29)}
\EE
(cf. (1.6)) reads:
\BE
E-2{L \over \Lambda} \left( {8 \over {n+8}}\right)^{1/2} {1 \over \sqrt{E}} =
0\,.
\EE

We see the ``repel'' from the classical constraint $E=0$.
Solution of (5-29)
\BE
\bar E=\left[{32 \over{n+8}}{{\Lambda^2}\over {L^2}}\right]^{1/3}
\label{(5-28)}
\EE
is the renormalized ``vacuum'' value of $E$; $E \ll \Lambda$,
i.e. small $E$ approximation \req{(5-24)}  requires $\Lambda\cdot L^2\gg1$.

\section{Conclusion}
More work and better understanding are necessary in the future. It would be
interesting to consider a more realistic theory instead of toy  models. Our
statement: in the EA language, the presence of a constraint shows up in the
quantum repel from the mass shell; this is demonstrated in~\req{(4-18)}
and~\req{(5-27)}. Concrete formulae for this phenomenon evidently depend on
the choice of the gauge-fixing condition, which is not a surprise.
Constraints are intimately connected with the gauge sector; all effects
studied in this report are of longitudinal nature.  However this does not
mean that the results are unphysical; it is well known after all that in the
relativistic gauges of nonabelian theories the very separation of the
longitudinal and transverse sectors is problematic~\cite{21}. A choice of the
gauge fixing means physically the choice of gauge-invariant observables.  If
gauge transformations are zero at infinity, they do not influence asymptotic
observables. But global, `large', gauge transformations cannot help influence
physical predictions of the theory. And even if we, following the VDWEA
approach, postulate Landau-\hbox{De$\,$Witt} gauge, the way of averaging over
residual
gauge modes in essential. We already discussed it in Sec.~4.

Let us sketch here the way of averaging alternative to the choice of the
noncausal contour in~\req{(4-15)} or  \req{(5-21)}.  We may define functional
integral in~\req{(3-11)} as a product of the ordinary functional integration
over fields vanishing at infinity and the integration over residual gauge
modes~\req{(3-18)}. Because of~\req{(3-17)} this will give logarithmic
divergency of the EA when  $S_{,B}\to0$. We shall not dwell on this approach
here, but in order to illustrate the idea we present symbolically the expected
EA for gravity:

\BE
\Gamma(\bar g)=S_E+{\rm const}\cdot\log\left\{
\det_{ij}\left[\int V_{ij}^{(\mu\nu)}(R_{\mu\nu}-{1 \over 2} g_{\mu\nu}R)
\sqrt{-g}\,d^4x\right]\right\}
\label{(6-1)}
\EE

\noi
where $S_E$ is the Einstein Action, $V_{ij}^{(\mu\nu)}$ are some standard
tensors constructed from the functional Cristoffel symbols, gauge Killings
$K_\b^{(\mu\nu)}$, and combinations $\xi^\alpha_{0i}\xi^\b_{0j}$ of all
solutions to Eq.~\req{(3-10')}; variations of the space-time metric
$\eta^{(\mu\nu)}_{\xi_{0i}}$ \req{(3-6)'} are normalized to unity with the
functional metric~\req{(3-4)}. It is easily seen that dynamical
equations~\req{(1-6)} given by the variations of~\req{(6-1)} are rather
Machian: they forbid empty Einstein spaces which obey $R_{\mu\nu}-{1\over 2}
g_{\mu\nu}R=0$.  Although the action~\req{(6-1)} differs drastically from the
conventional GR dynamics, there are forcible grounds to conclude that it does
not come into contradictions with the local observations. We shall touch this
problem elsewhere.

The question arises: what is the `empty' or `nonempty' space?  Why do we
define external sources as a variation of the classical Action ($\delta
S/\delta\vF^A$) and not of the EA ($\delta\Gamma/\delta\vF^A$) ?  A possible
answer is that quantum corrections considered in this report are obligatory
nonlocal (as in~\req{(6-1)}) (they just look as local in~\req{(5-27)} because
we considered homogeneous background).  Thus it is physically natural to
define external sources as nonzero variations of the local Action $S$ (let
$S$ include all quantum ultraviolet renormalizations), whereas
$\delta\Gamma^{\rm(non\ local)}/\delta\vF^A$=0 \req{(1-6)} is a
self-consistency condition to determine external currents.

\medskip

Describing the Universe with a modified Einstein equations was in a trend of
thought of Academician Markov. The integral formulation of Einstein
equations \cite{22,23}
\BE
g_{ik}(x)=\kappa\int\cD_{ik}^{\alpha\b}(x,y)T_{\alpha\b}(y)
\sqrt{-g}\,d^4y
\label{(6-2)}
\EE

\noi
was also a topic of his interest. This field is  intimately connected
with the ideas of the present report. Let us consider instead of~\req{(6-2)}
the integral equation for a small ``residual'' gauge variations (3.7)
of the metric (here $\xi = \xi_0$ is the solution of (3.13)):
\BE
\delta^{(\xi)}g_{ik}=
\kappa\int\cD_{ik}^{\alpha\b}\delta^{(\xi)}T_{\alpha\b}
\sqrt{-g}\,d^4y\,,
\label{(6-3)}
\EE

\noi where $\kappa\delta^{(\xi)}T_{\alpha\b}=\vec F_{\alpha\b}^{\g\delta}
\delta^{(\xi)}g_{\g\delta}$ are the gauge variations of the external current,
$\vec F$ is the VDW differential operator defined in~\req{(3-12)}, $\cD={\vec
F}^{-1}$.  Eq.~\req{(6-3)} puts rather strong demands upon the background
metric $\bar g^{ik}$;
as well as~\req{(6-2)}, it forbids empty space-times. \req{(6-3)}
means that residual gauge variations of the metric are totally created by the
corresponding variations of the sources, there are
no `free' $\delta g_{ik}^{(\xi)}$ and hence no Faddeev--Popov zero modes in
the functional integral (3.14).  We may postulate that only the
backgrounds, where~\req{(6-3)} is valid, are realized in Nature.  Supposedely
quantum functional averaging over residual gauge modes considered in this
report is a dynamical way to the selection rule~\req{(6-3)}.  In \cite{23}
Mal'tsev and Markov proposed a scalar version of the integral form~\req{(6-2)}.
It would be interesting to investigate for this simplified model the quantum
approach of this report and its connection with the infinitesimal integral
form (6.3).

\subsection*{Acknowledgement}
One of the authors (B. A.) is grateful for many useful discussions to A.
Barvinsky, I. Batalin, M.~Iofa, D. Kirzhnits, V. Man'ko, A. Mironov, G.
Ryazanov, M. Soloviev, I.~Tyutin, G.~Vilkovisky and B.~Voronov.


\begin{thebibliography}{99}
\addtolength{\baselineskip}{-.21\baselineskip}
\parindent=0pt
\parskip=-5pt
\bibitem{1}  A.O.~Barvinsky, Phys.\ Rep.\ {\bf 230} (1993) 237.

\bibitem{2} M.B.~Mensky, Class.\ Quantum Grav.\ {\bf 7} (1990) 2317.

\bibitem{3}S.~Coleman and E.~Weinberg, Phys.\ Rev.\ {\bf D7} (1973) 1888.

\bibitem{4}B.L.~Altshuler, A.M.~Boyarsky, and A.Yu.~Neronov, Gravitation and
Cosmology, {\bf 1} (1995) 191.

\bibitem{5} L.V.~Prokhorov and S.V.~Shabanov, Phys.\ Lett.
\ {\bf B216} (1989) 341; Usp.~Fiz.~Nauk,\
          {\bf 161} (1991) 213; Int.\ Journ.\ Mod.\ Phys.\ {\bf A7} (1992)
7815.
\bibitem{6}  S.K.~Blau, Ann.\ Phys.\ (USA)\  {\bf 205} (1991) 392.
\bibitem{7}  P.~Senjanovic, Ann.\ Phys.\ (USA)\ {\bf 100} (1976) 227.
\bibitem{8}  P.A.M.~Dirac, Rev.\ Mod.\ Phys.\ {\bf 17} (1945) 195;
       B.R.~Holstein, Amer.\ J.\ Phys.\ {\bf 56} (1988) 425.
\bibitem{9}  F.J.~Dyson. J.\ Math.\ Phys.\ {\bf 3} (1962) 140, 1191,1199;
       E.~Brezin, G.~Itzykson, G.~Parisi, and J.B.~Zuber.
       Comm.\ Math.\ Phys.\ {\bf 59} (1978) 35.
\bibitem{10}  C.~Teitelboim, Phys.\ Rev.\ {\bf D25} (1982) 3159;
       {\bf D28} (1983) 298, 310.
\bibitem{11}  J.J.~Halliwell, Phys.\ Rev.\ {\bf D38} (1988) 2468.
\bibitem{12}  E.I.~Guendelman, Int.\ J.\ Mod.\ Phys.\ {\bf A5} (1990) 2783.
\bibitem{13}  E.S.~Fradkin and A.A.~Tseytlin, Nucl.\ Phys.\ {\bf B234} (1984)
509.
\bibitem{14}  Yu.A.~Golfand, in ``Quantum Field theory and Quantum Statistics",
   ed.  I.A.~Batalin, C.J.~Ishaw,  and G.A.~Vilkovisky.
         Adam Hilger Bristol. 1987, pp.~223-244.
\bibitem{15}  H.J.~Borchers, Nuovo\ Cim.\ {\bf 25} (1960) 279.
\bibitem{16}  B.S.~DeWitt, Phys.\ Rev.\ {\bf 162} (1967) 1195;
      in ``General  Relativity. An Einstein Centenary Survey",
            ed.\ S.W.~Hawking and W.~Israel, Cambridge University Press, 1979.

\bibitem{17}  G.A.~Vilkovisky, Nucl.\ Phys.\ {\bf B234} (1984) 125;
         Class.\ Quantum Grav.\ {\bf 9} (1992) 895; in ``Quantum theory of
Gravity",
         ed.: S.M.~Christensen, Adam Hilger, Bristol,1984.

\bibitem{18}J.F.~Barbero and J.~Perez-Mercader, Phys.\ Rev.\ {\bf D48} (1993)
3663;
           R.~Kantowski and C.~Marzban, Phys.\ Rev.\ {\bf D46} (1992),5449;
           H.T.~Cho, Phys.\ Rev.\ {\bf D43} (1991) 1859;
           S.R.~Huggins, G.~Kunstatter, H.P.~Leivo, and D.J.~Toms,
           Nucl.\ Phys.\ {\bf  B301} (1988) 627.

\bibitem{19}R.J.~Epp, G.~Kunstatter, and D.J.~Toms, Phys.\ Rev.\ {\bf D47}
(1993) 2474;\\
J.~Balakrishnan and D.J.~Toms,  Phys.\ Rev.\ {\bf D46} (1992) 4413.

\bibitem{20}D.M.~Gitman and I.V.~Tyutin,
{\it Quantization of Fields with Constraints\/}, Springer--Verlag,
Berlin, 1990.

\bibitem{21}M.A.~Soloviev, Teor.\ Mat.\ Fiz.\ {\bf 78} (1989) 163.

\bibitem{22}B.L.~Altshuler, ZhETF\ {\bf 51} (1966) 1143
(Sov.\ Phys.\ JETP\ {\bf 24} (1967) 766);
Int.\ J.\ Theoret.\ Phys.\ {\bf 24} (1985) 99;\\
D.~Lynden-Bell, Month.\ Not.\ Royal Astron.\ Soc.\ {\bf 135} (1967) 413;\\
D.W.~Sciama, P.C.~Waylen, and R.C.~Gilman,
Phys.\ Rev.\ {\bf D2} (1970) 1400.

\bibitem{23}V.K.~Mal'tsev and M.A.~Markov, Trudy FIAN\ {\bf 96} (1977) 11.

\end{thebibliography}
\end{document}